\documentclass[manuscript]{aastex}
\pdfoutput=1

\newcommand{\rahms}[4]{$#1^{\rm h}#2^{\rm m}#3\mbox{$^{\rm s}\mskip-7.6mu.\,$}#4$} 
\newcommand{\decdms}[4]{$#1^{\circ}#2'#3\mbox{$''\mskip-7.6mu.\,$}#4$} 
\newcommand{\msec}[2]{$#1\mbox{$''\mskip-7.6mu.\,$}#2$}

\newcommand{\hii}{\mbox{\ion{H}{2}}}

\shorttitle{Compact Radio Source in W3(OH)}
\shortauthors{Dzib et al.}

\begin{document}

\title{The compact, time-variable radio source
projected inside W3(OH): Evidence for a Photoevaporated Disk?}

\author{Sergio A. Dzib\altaffilmark{1}, 
Carolina Rodr\'\i guez-Garza\altaffilmark{1},
Luis F. Rodr\'\i guez\altaffilmark{1,2},
Stan E. Kurtz\altaffilmark{1},
Laurent Loinard\altaffilmark{1},
Luis A. Zapata\altaffilmark{1},
and
Susana Lizano\altaffilmark{1}}

\email{s.dzib@crya.unam.mx}

\altaffiltext{1}{Centro de Radiostronom\'ia y Astrof\'isica, Universidad Nacional Aut\'onoma de Mexico,
Morelia 58089, Mexico}
\altaffiltext{2}{Astronomy Department, Faculty of Science, King Abdulaziz University, P.O.
Box 80203, Jeddah 21589, Saudi Arabia}

\begin{abstract}
We present new Karl G. Jansky Very Large Array observations of the compact ($\sim$ \msec{0}{05}), 
time-variable radio source
projected near the center of the ultracompact HII region W3(OH). 
The analysis of our new data as well as of VLA archival observations
confirms the variability of the source on timescales of years
and for a given epoch indicates
a spectral index of $\alpha = 1.3 \pm 0.3$ ($S_\nu \propto \nu^\alpha$).
This spectral index and the brightness temperature of
the source ($\sim$6,500 K)
suggest that we are most probably detecting partially optically thick
free-free radiation. The radio source is probably associated with the ionizing star of
W3(OH), but an interpretation in terms of an ionized
stellar wind fails because the detected flux densities are
orders of magnitude larger than expected. 
We discuss several scenarios and tentatively propose that the
radio emission could arise in a static ionized atmosphere around
a fossil photoevaporated disk.
\end{abstract}


\section{Introduction}

Recently formed massive stars produce large amounts of UV photons
that ionize the natal material around them, forming \hii~ regions
of different morphologies and sizes.
Ultracompact \hii~ (UCHII) regions are small (diameters no more than $\sim$0.1 pc)
and dense (electron densities of at least $10^4$ cm$^3$)
volumes of ionized gas produced by one or several
massive stars (Wood \& Churchwell 1989). W3(OH) is a limb-brightened UCHII
region with a shell morphology (Dreher \& Welch 1981) that is known to be expanding
at a velocity of $\sim$3-5 km s$^{-1}$ (Kawamura \&
Masson 1998). It is located at a distance of 2.04 kpc (Hachisuka 
et al.\ 2006) and the total luminosity of the system
is $7.1 \times 10^4~L_\odot$
(Hirsch et al. 2012). The region is heavily
obscured (A$_V\sim$75; Feigelson \& Townsley 2008) 
and there is little direct information on the nature of
the ionizing star.

Kawamura \& Masson (1998), based on the residual image obtained 
subtracting maps of different epochs, reported a compact, time
evolving (over a scale of several years) source projected near the center of
W3(OH) and suggested that it is probably related to
the central star. Remarkably, no further research has been published since
then on this interesting radio source.
Here, we report new radio observations of the W3(OH)
region made with the Karl G. Jansky Very Large Array
(VLA) of the NRAO\footnote{The National Radio
Astronomy Observatory is operated by Associated Universities
Inc. under cooperative agreement with the National Science Foundation.}. We 
also used VLA archive observations to complement
our study. Our results provide for the first time a direct radio
detection and a determination of the parameters of this radio source,
possibly associated with the star that ionizes the
W3(OH) region.

\section{Radio Observations}

The observations were made in the Ka band (26.5 -- 40 GHz) with the VLA 
centered at a frequency of 32.96 GHz and
with a total bandwith of 2 GHz. The observations were made
on 2012 October 13, with the VLA in the A configuration.

At the beginning of the observations we integrated on 
the standard flux calibrator 3C 48 for $\sim$3 minutes.
This source was also used as the bandpass calibrator. We then
spent one minute on the phase calibrator
J0244+6228 followed by five minutes on the target; this
cycle was repeated until one hour was completed. 
The angular distance between the phase 
calibrator and the target is 2.17 degrees. 
Referenced pointing scans at the lower frequency of 9.0 GHz were
performed before the beginning of the observation of the flux calibrator
and before the start of the phase calibrator--target cycle.
These scans are required to assure that the absolute pointing
of the antennas is accurate to $5''$ or better. 

The data were edited, calibrated and imaged in the standard
fashion using the Common Astronomy Software Applications
package (CASA). After the initial calibration, the visibilities
were self-calibrated and imaged with a pixel size of 0.01 arcsecond. The
weighting scheme used was intermediate between
natural and uniform (WEIGHTING='briggs' with ROBUST=0.0 in CASA). 
To minimize the extended emission
from the \hii~ region we removed the short spacings generated by
baselines below 10 km, suppressing angular structures
larger than $\sim$ \msec{0}{2}. The uv range was chosen to provide an optimal compromise between
supressing extended emission and maintaining signal-to-noise and image quality in
the resulting maps. The rms noise level of an image processed in this manner
depends on the position considered. In the central part
of the final images it was around 50 $\mu$Jy beam$^{-1}$, while
in the immediate surroundings of the compact source
it reached about 380 $\mu$Jy beam$^{-1}$.

\subsection{Position and Angular Size of the Compact Radio Source}

A compact radio source was clearly detected at a position of
$\alpha$(2000)=\rahms{02}{27}{03}{867}, $\delta$(2000)=\decdms{61}{52}{24}{89}~
(see Figure 1). The source is projected well inside W3(OH) and its
total integrated flux density at 32.96 GHz is 14.4$\pm$1.0 mJy with a peak of
6.0$\pm$0.4 mJy beam$^{-1}$. Its deconvolved
angular size is \msec{0}{05} $\pm$ \msec{0}{01}, implying a physical
size of $100\pm20$ AU and a peak brightness
temperature of $T_B = 6,500 \pm 2,600$ K. This brightness temperature
suggests that we are observing optically-thick photoionized gas.
Assuming that the brightness temperature equals the electron temperature
of the gas and using the usual formulation for the parameters of a
homogeneous HII region
(Schraml \& Mezger 1969) we can set lower limits to the electron density, $n_e \geq
1.9 \times 10^6~cm^{-3}$, ionized mass, $M_{HII} \geq 9.5 \times 10^{-6}~M_\odot$,
and ionizing photon rate required by the compact source, $N_i \geq 7.8 \times 10^{45}~s^{-1}$. 
This photon rate is equivalent to that provided by 
a B1 or earlier ZAMS star. There is marginal evidence in the image of
a very faint, extended halo but observations of higher signal-to-noise
ratio are needed to test its reality.

\begin{figure*}[!t]
  \centerline{\includegraphics[height=0.8\textwidth,angle=0]{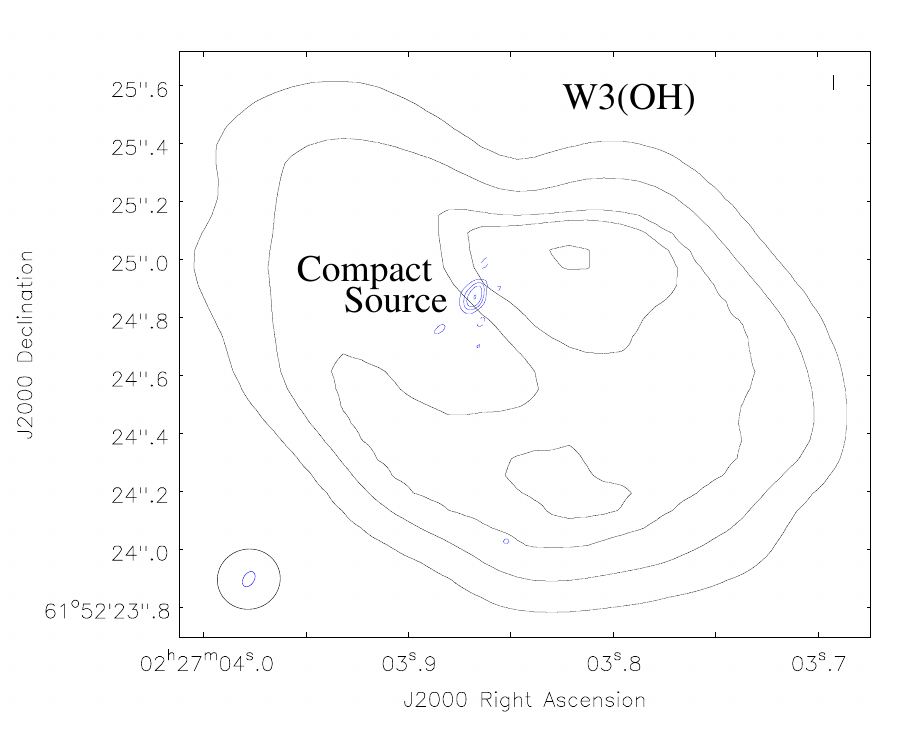}}
   \caption{Image of the I Stokes parameter of the compact radio source
   in W3(OH), given in blue contours, as detected
   in our Ka band (32.96 GHz) observations. The contours are at -3, 3, 6, 9, and 15
   times the 1-$\sigma$ noise level of 380 $\mu$Jy beam$^{-1}$.
   The black contours delineate the W3(OH) region as observed at a frequency of 8.4 GHz.
   The contours are at 320, 1000, 1500, 1600, and 1750 times the 1-$\sigma$ noise
   level of this image (22 $\mu$Jy beam$^{-1}$). The half power contours of the
   32.96 ($0\rlap.{''}057 \times 0\rlap.{''}037; PA = 148^\circ$) and 8.4 GHz synthesized beams 
   ($0\rlap.{''}218 \times 0\rlap.{''}207; PA = 105^\circ$) are shown in the bottom left corner.}
    \label{fig:ka+c}
    \end{figure*}

In Figure 1 we also show the extent of W3(OH) as observed at 8.4 GHz. These data were
taken from the VLA archive and are the average of observations made in the A configuration during
1995 July 04 (under project AR341) and 1996 November 15 (under project AR363). We will
refer to these averaged 8.4 GHz data as having a mean epoch of 1996.19. 
These archive data have been discussed and published by Wilner et al. (1999).

We have also compared the position of the compact radio source with that of
the UCHII region W3(OH), as traced in the 890 $\mu$m data taken with the
Submillimeter Array (SMA)\footnote{The Submillimeter Array (SMA) is a joint project 
between the Smithsonian Astrophysical Observatory and the Academia Sinica Institute 
of Astronomy and Astrophysics and is funded by the Smithsonian Institution and the 
Academia Sinica.}
and discussed by
Zapata et al. (2011). The total emission of W3(OH) at this wavelength is
determined to be 1.8$\pm$0.1 Jy and the image is shown in Figure 2. The compact radio
source falls near the centroid of the 890 $\mu$m emission.
{There is an offset of $\sim$\msec{0}{3} between the centroids of the compact
radio source and the SMA submillimeter emission. However, the positional
accuracy of the submillimeter emission is $\sim$\msec{0}{2} (Cowie et al. 2009)
and we cannot attribute significance to this offset.}

\begin{figure}[!t]
  \centerline{\includegraphics[height=0.45\textwidth,angle=0]{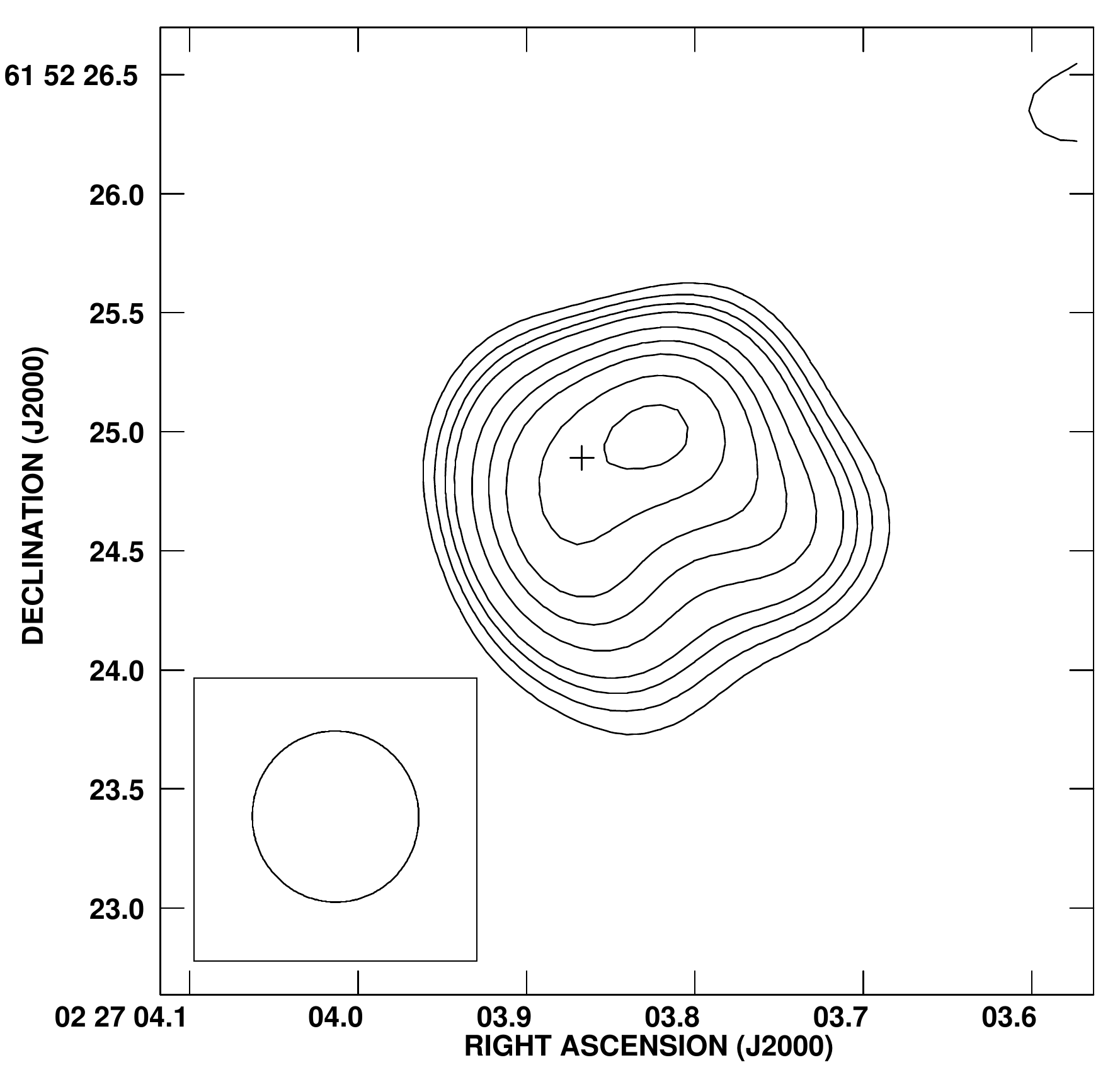}}
 \caption{SMA 890 $\mu$m continuum emission from
W3(OH). The contours are at -3, 3, 4, 5, 6, 8, 10, 12, 15, and 18
times the 1-$\sigma$ noise level of 36 mJy beam$^{-1}$.
The half power contour of the synthesized beam
($0\rlap.{''}72 \times 0\rlap.{''}70; PA = 5^\circ$) 
is shown in the bottom left corner. The position of the compact
radio source is marked with a cross.}
 \label{fig:890mu}
  \end{figure}

\subsection{{Detection of the H58$\alpha$ line}}

The Ka band observations were made centered at 32.96
GHz with a total bandwidth of 2 GHz distributed
in 1024 individual channels of 1.95 MHz each (equivalent to an average velocity width of
17.8 km s$^{-1}$).
This frequency coverage includes the H58$\alpha$ radio recombination line,
that has a rest frequency of 32.85220 GHz. We analyzed the part of the
spectrum where this line was expected and a feature was detected (Figure 3).
This line was least-squares fitted with a Gaussian profile, obtaining
a peak line flux density of $S_L$ = 4.6$\pm$1.0 mJy, a full width at half maximum of
$\Delta v$ = 52.9$\pm$13.1 km s$^{-1}$, and an LSR radial velocity of $v_{LSR}$ = 
$-$58.5$\pm$5.6 km s$^{-1}$.
The analysis was made in the spectral cube obtained 
without short spacings and in the solid angle containing the compact
radio source, with a continuum flux density of
$S_C$ = 14.4$\pm$1.0 mJy. 

\begin{figure}[!t]
  \centerline{\includegraphics[height=0.50\textwidth,angle=0]{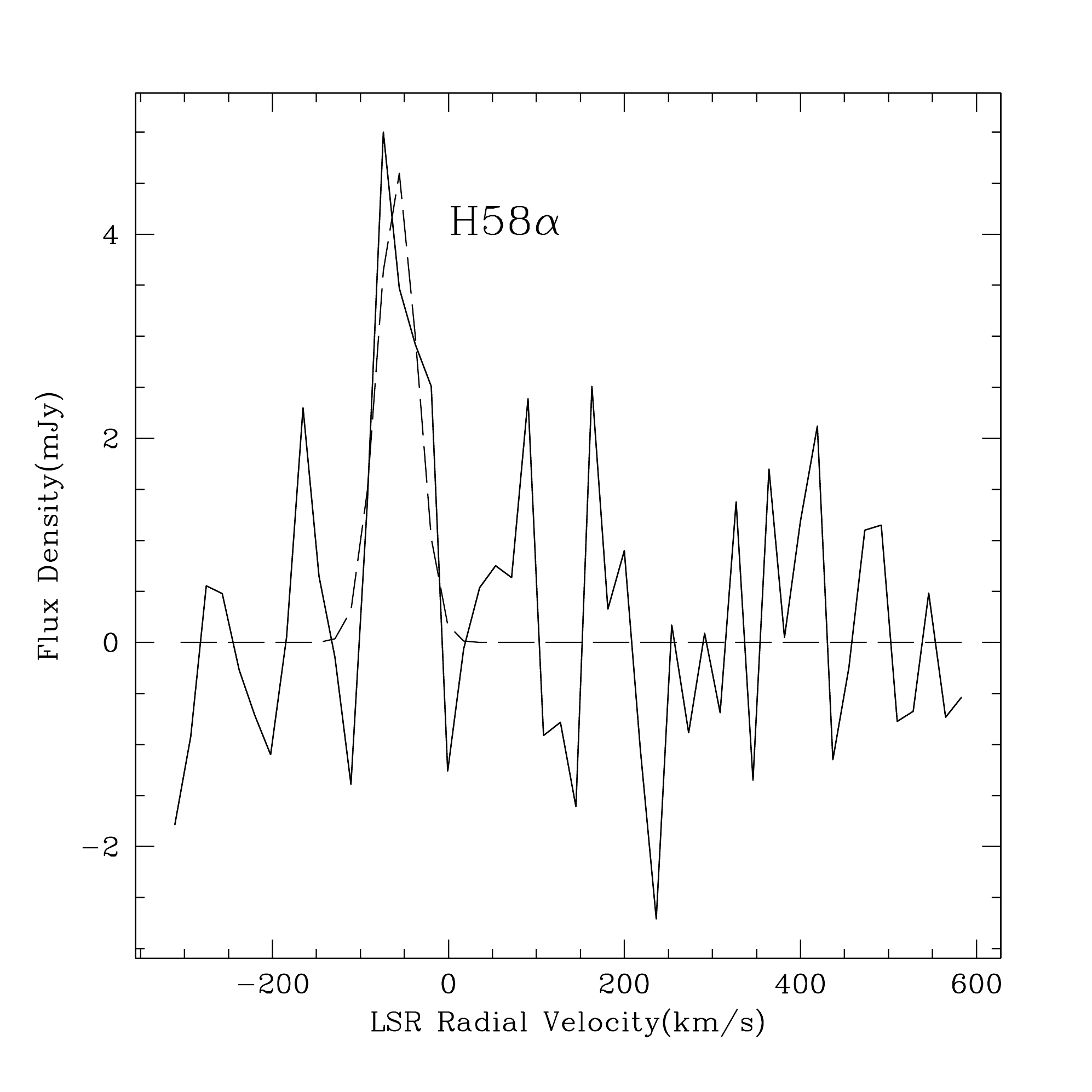}}
   \caption{H58$\alpha$ emission from the compact radio source in W3(OH).
   The dashed line is the least-squares fit to the spectrum.}
    \label{fig:hh58a}
  \end{figure}

We believe that this line corresponds to emission from the compact
radio source by two reasons. The first is that the $v_{LSR}$ of the
line falls in the range of LSR radial velocities reported for radio
recombination lines originating in the UCHII region W3(OH) ($-$63.5 to $-$50.3 km s$^{-1}$;
Sams et al. 1996). The second reason is that the observed flux density for the line
is in agreement with the expected value. Using the formulation of
Rodr\'\i guez et al. (2009) for radio recombination lines from an
ionized wind in local thermodynamic equilibrium, we expect a value of $S_L$ = 3.9 mJy, consistent with the
observed value of 4.6$\pm$1.0 mJy.

\subsection{Spectral Index}

We used the VLA archive deep observations of W3(OH),
previously reported by Kawamura \& Masson (1998, see
their paper for a detailed description of the observations)
made in the years 1986, 1990 and 1995.
We removed short spacings generated by baselines
below 8 km. The source is detected in the three epochs
(see Table 1). It is strongly variable on a timescale of years, showing 
minima in 1990 and 1995 and rising again to its 1986 levels in our
2012 observations. Even when the observations are made at different
frequencies, it is clear that the flux densities of 1990 and 1995 represent
minima with respect to the other epochs after correcting for
the spectral index determined from the 1986 observations (see below).  
Its flux densities for 1986, 1990 and 1995 are consistent with the negative values
reported by Kawamura \& Masson (1998) in their residual
difference maps. 

In the epoch 1986.38 the source was detected at the two observed
frequencies, allowing us to determine a spectral index of
$\alpha = 1.3\pm0.3$ ($S_\nu \propto \nu^\alpha$) for this epoch. This spectral 
index is similar to 
the one reported for the compact radio source at the center of the \hii~ region
NGC 6334E ($\alpha = 1.0\pm0.7$), that Carral et al.\ (2002) interpreted as produced
by a stellar ionized wind. We also imaged the Stokes V parameter in all the epochs
and detected no circular polarization at levels of $\sim$1 -- 10\% (4-$\sigma$
upper limits). The long timescale variability, along with
the lack of polarization, the positive spectral index, the possible
presence of recombination line emission, and the brightness temperature
of $\sim10^4$ K favor optically thick free-free as the emission process.
Gyrosynchrotron radiation from a magnetically active young star of low mass
would likely be variable on short timescales (hours),
show some degree of circular polarization, have a negative spectral index and
a high ($\sim10^7$ K) brightness temperature 
(e.g. Dzib et al. 2011; Torres et al. 2012).

\begin{table*}[htb]
\small
\begin{center}
\caption{Position, flux density and observed frequency of the compact radio source in the different epochs.}
\begin{tabular}{lcccccc}\hline\hline
  & $\alpha$(J2000.0) & $\sigma_{\alpha}$ & $\delta$(J2000.0) & $\sigma_\delta$ & $F_\nu$ & $\nu$ \\
  Epoch   & $2^{\rm h}27^{{\rm m}}$& (seconds) & $+61^{\circ}52'$  & (arcseconds)  & (mJy)   & (GHz)\\
	 \hline
       1986.38 &    03.8709 &  0.0002     &     24.919     &    0.003     &    09.1$\pm$0.5 &14.99\\
       1986.38 &    03.8707 &  0.0002     &     24.906     &    0.003     &    15.6$\pm$1.3 &22.46\\
       1990.19 &    03.8770 &  0.0004     &     24.953     &    0.005     &    03.8$\pm$0.4 &14.94\\
       1995.48 &    03.8707 &  0.0003     &     24.890     &    0.005     &    02.6$\pm$0.2 &14.94\\
       2012.78 &    03.8670 &  0.0001     &     24.887     &    0.002     &    14.4$\pm$1.0 &32.92\\
   \hline\hline
     \label{tab:4astrometry}
  \end{tabular}
	\end{center}
     \end{table*}

\section{Discussion}

\subsection{Proper Motions}

In our analysis of the available data we have used the improved positions
for the phase calibrators available in the VLA Calibrator Manual.
This procedure permits absolute systematic errors in the
range of $\sim0\rlap.{''}01 - 0\rlap.{''}02$ (e.g. G\'omez et al. 2005).
With the positions presented in Table 1, 
we determine the proper motion of the compact radio source to be \footnote{We do not 
include epoch 1990.19
in our proper motion analysis
because its position does not follow the trend presented by the others, with a difference
of $\sim$ \msec{0}{05}.}:

\begin{eqnarray}
\mu_\alpha\cos \delta  & = &  -2.3 \pm 0.6 \mbox{~mas~yr$^{-1}$}\nonumber\\
\mu_\delta  & = &  -1.1 \pm 0.7 \mbox{~mas~yr$^{-1}$.} \nonumber
\end{eqnarray}

Systematic contributions of \msec{0}{010} and \msec{0}{011}
in the $\alpha$ and $\delta$ positions, respectively,
were added in quadrature to the positional errors obtained from a Gaussian fit
(task IMFIT in CASA), in order to obtain a $\chi^2$ of 1.
The positions as a function of the epoch are presented in Figure 4.

\begin{figure*}[!t]
  \centerline{\includegraphics[height=0.55\textwidth,angle=0]{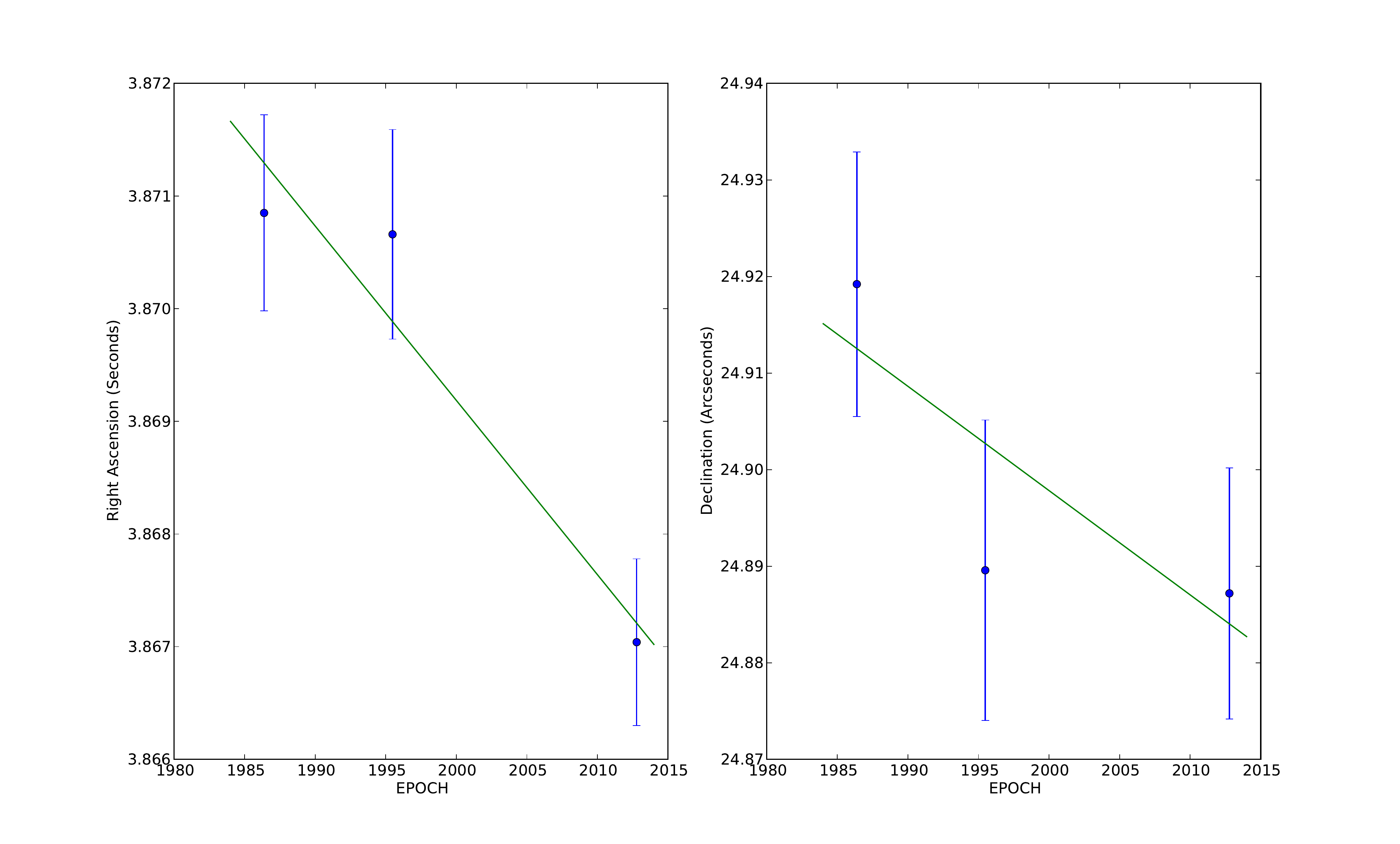}}
  \caption{Positions of the compact radio source as a function of epoch.
  The solid lines are the least squares best fits to the positions.}
\label{fig:propermotions}
  \end{figure*}

Does this proper motion have any significance with respect to the nature of
the source? To estimate the characteristic proper motion of the region
we compared our 2012 observations with those made at 8.4 GHz for
epoch 1996.19 for the TW sources A and C (Turner \& Welch 1984; Reid et al. 1995), located about
$7''$ to the east of W3(OH). As can be seen in Figure 5, there is a small
displacement to the west for the more recent data. TW-A is extended in the east-west
direction and not adequate for a proper motion determination that
requires sources as compact as possible. Fortunately, TW-C is very compact
and for it
we estimate a proper motion of

\begin{eqnarray}
\mu_\alpha\cos \delta  & = &  -2.7 \pm 0.3 \mbox{~mas~yr$^{-1}$}\nonumber\\
\mu_\delta  & = &  -0.3 \pm 0.3 \mbox{~mas~yr$^{-1}$.} \nonumber
\end{eqnarray}

These proper motions are consistent with those found
for the compact radio source and suggest that we are simply observing
the secular proper motions of the region.
From VLBA observations, Xu et al. (2006) measure a mean proper motion of

\begin{eqnarray}
\mu_\alpha\cos \delta  & = &  -1.20 \pm 0.02 \mbox{~mas~yr$^{-1}$}\nonumber\\
\mu_\delta  & = &  -0.15 \pm 0.01 \mbox{~mas~yr$^{-1}$.} \nonumber
\end{eqnarray}

\noindent for methanol masers associated with W3(OH).
To obtain the expected
galactic proper motions we used the galactic rotation
model of Brand \& Blitz (1993) and the velocity
of the Sun with respect to the local standard of rest
from Sch{\"o}nrich et al. (2010). We obtain:

\begin{eqnarray}
\mu_\alpha\cos \delta  & = &  -0.8 \pm 1.0 \mbox{~mas~yr$^{-1}$}\nonumber\\
\mu_\delta  & = &  -0.4 \pm 1.0 \mbox{~mas~yr$^{-1}$,} \nonumber
\end{eqnarray}

\noindent where the associated error range is derived assuming that the
observed region could have peculiar velocities of up to $\pm$10 km s$^{-1}$
(Stark \& Brand 1989).
We then conclude that the observed proper motions are roughly consistent with
those expected for a source at the position of W3(OH). 

\begin{figure*}[!t]
  \centerline{\includegraphics[height=0.5\textwidth,angle=0]{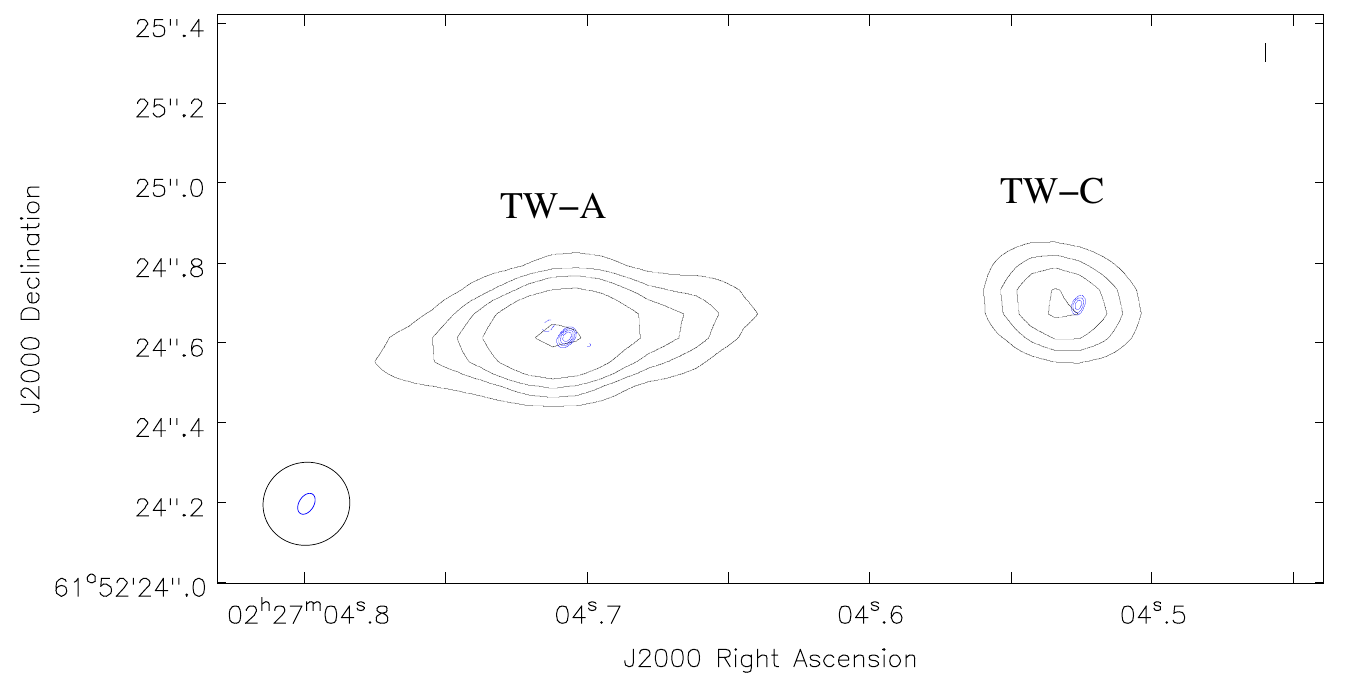}}
     \caption{Image of the I Stokes parameter of the compact radio sources TW-A (left)
     and TW-C (right),
given in blue contours, as detected
in our Ka band (32.96 GHz) observations of epoch 2012.78. The contours are at -4, 4, 5, 6 and 7
times the 1-$\sigma$ noise level of 52 $\mu$Jy beam$^{-1}$.
The black contours delineate the 8.4 GHz emission from epoch 1996.19.
The contours are at -4, 4, 8, 12, 18, 36 times the 1-$\sigma$ noise
level of this region of the image (15 $\mu$Jy beam$^{-1}$). Note the small
displacement in position for the peak emission between the
two epochs. The half power contours of the
32.96 ($0\rlap.{''}057 \times 0\rlap.{''}037; PA = 148^\circ$) and 8.4 GHz synthesized beams
($0\rlap.{''}218 \times 0\rlap.{''}207; PA = 105^\circ$) are shown in the bottom left corner.}
	 \label{fig:tw-ka-x}
   \end{figure*}

\subsection{Counterparts at Other Wavelengths}

At the epoch of the Kawamura \& Masson (1998) indirect detection of the
compact radio source there were no known counterparts at other wavelengths.
More recent studies indicate the presence of 
counterparts at several wavelengths.
A Class 0/I infrared young stellar object was detected by 
Rivera-Ingraham et al.\ (2011) at a position 
$\alpha$(2000)=\rahms{02}{27}{03}{86}, 
$\delta$(2000)=\decdms{61}{52}{25}{32}.
As they used Spitzer data we assume that their 
astrometric errors are between \msec{0}{35}
and \msec{1}{0} (Fadda et al.\ 2006). Thus, within positional
errors, this infrared source coincides with the radio source.

We also searched in the 2MASS catalog (Cutri et al.\ 2003) and found a 
source, 2MASS 02270391+6152255, at an angular distance of \msec{0}{5}
from the position of the radio source. Since the positional accuracy
of the 2MASS survey, with respect to the International Celestial Reference System, 
is \msec{0}{5} (Skrutskie et
al.\ 2006), we also consider this 2MASS source a counterpart to the compact
radio source. {However, since the star is enshrouded in an optically-thick
shell ($\tau\sim\,$2.8 at 37.1 microns, Hirsch et al. 2012) the emission observed
with Spitzer and 2MASS is probably coming from the outer edge of this dust shell,
at thousand to tens of thousands of AU from the star (Stecklum et al. 2002, Hirsch
et al. 2012).}

Feigelson \& Townsley (2008) detected a hard X-ray source
coincident within error with the compact radio source.
In their words:
{\it The young massive star ionizing W3(OH) is clearly
detected in our Chandra observation, at
$\alpha$(2000)=\rahms{02}{27}{03}{84},
$\delta$(2000)=\decdms{61}{52}{24}{9}.
It is a surprisingly hard X-ray source;
this hard emission allows it to be seen through a large
absorbing column (A$_{V}\sim$ 75 mag) inferred from the
soft X-ray absorption.}
Their positional error is $\sim$ \msec{0}{4}.

Given that these sources detected in different surveys
and at different wavelengths coincide within their
positional errors with the compact radio source, we suggest 
that the latter is related with a very young massive stellar object
projected inside the W3(OH) ultracompact \hii, possibly its exciting star.

\subsection{Similar Sources in the Literature}

A previous detection of compact radio sources near the center
of the HII regions
NGC 6334E and NGC 6334A (both with shell morphologies) was 
reported by Carral et  al.\ (2002). These
compact sources were suggested to be associated with the 
exciting stars of the HII regions.

\section{Interpretation}
What is the nature of the compact radio source? Its positional coincidence 
with an embedded star, brightness temperature, spectral index and lack
of circular polarization seem to favor a free-free emitting ionized stellar wind.
There are, however, serious problems with this interpretation.  
Dzib et al.\ (2013) have tabulated the expected flux densities for
ZAMS stars of different classes. The luminosity of W3(OH) ($7.1 \times 10^4~L_\odot$;
Hirsch et al. 2012)
and the rate of ionizing photons required to maintain its
free-free emission of $\sim$2 Jy in the optically-thin regime
($\sim 9 \times 10^{47}$ s$^{-1}$) can be provided by a B0V star,
according to the tabulation of Dzib et al. (2013).
For a B0V star located at a distance
of 2.04 kpc, the expected flux density of its associated
stellar wind at 22.64 GHz is only 0.006 mJy,
about three orders of magnitude below the observed values.
We note that the compact sources detected near the center of the
HII regions NGC 6334(A) and NGC 6334(E) (Carral et al. 2002)
are much brighter (by about two orders of magnitude) than the expected
values for the O7.5 stars needed to ionize them.
Furthermore, the compact radio source in W3(OH) shows strong time variation while
the wind from a ZAMS star is expected to be steady.
Finally, the observed width of the H58$\alpha$ line suggests that
we are observing an HII region or a photoevaporated wind, since much
larger linewidths would be expected for a stellar wind.

Recent radio observations have shown time variability in HII regions 
on a timescale of years
(Franco-Hern\'andez \& Rodr\'\i guez 2004; 
Galv\'an-Madrid et al. 2008). Franco-Hern\'andez \& 
Rodr\'\i guez (2004) proposed that the time variability may be 
due to changes in the source of the ionizing radiation, though 
it may also be due to increased absorption in the rapidly evolving core of the nebula.
Peters et al. (2010a; 2010b), Galv\'an-Madrid et al. (2011) and Klassen et al. (2012a; 2012b) 
have discussed theoretical scenarios that may account for the HII region variability.
However, these observations and models address a situation in which the
whole of the HII region changes. This does not seem to be the case
for W3(OH), where the variation is restricted
to the compact component while the whole of the nebula does not 
vary significantly (Kawamura \& Masson 1998).

We tentatively propose that the compact radio source could be the result of
either a slow shell ejection by the B0 star or by the passage of a dense gas clump
that temporarily engulfs the star. 
In both cases we will have the dense gas close to the star that is
required to explain the free-free emission. These possibilities, however,
are rather \sl ad hoc\rm, and we explore the scenario of a photoevaporated disk,
a fossil of the star formation process, around the exciting star
of W3(OH). Using an expansion velocity of $\sim$4 km s$^{-1}$ for W3(OH)
(Kawamura \& Masson 1998) and a radius of 0.006 pc, W3(OH) has a kinematic age
of only 1,500 years and the accretion disk that presumably existed
around the star during its formation could still be present.

A compact ionized region around a young massive
star can be produced by
a photoevaporated disk (Hollenbach et al. 1994;
hereafter H94).
In the case of W3(OH) the compact source has a
size of the order of 100 AU. This size corresponds
to the gravitational radius $r_g = G M_*/a^2$ where
$M_*$ is the stellar mass and $a$ is the sound speed of the ionized gas. Beyond
this radius, the photoevaporated gas can escape from
the potential well of the star. Inside $r_g$ the gas
is trapped and will form a static atmosphere that
can produce the observed free-free emission. 

As discussed above,  the exciting source of W3(OH) has a luminosity
$L_* \sim 7.1 \times 10^4 L_\odot$. According to Vacca, Garmany \& Shull (1996), a 
ZAMS star with this luminosity
has a stellar mass, $M_* \sim 19.5~ M_\odot$, a stellar radius $R_* \sim 8.3~ R_\odot$,
and a rate of ionizing photons $\dot N_i \sim 1.44 \times 10^{48}~ {\rm s}^{-1}$. 
For this star,  the gravitational
radius is $r_g \sim 130$ AU. The static atmosphere has
an exponential electron number density profile $n_e(r,z) = n_0(r) \exp(-z^2/2 H^2)$,
as a function of height $z$ and radius $r$, where $n_0(r)$ is the electron density at the disk surface and
$H(r)=r_g(r/r_g)^{3/2}$ is the scale height. The number density $n_0(r)$ is given by eqn. (3.11) of
H94, and 
one can obtain the free-free optical depth of a face-on disk as a function of radius, 
assuming an electron temperature
of the ionized gas of $T_e \sim 6,500$ K. 

We find that the optical depth at 32.96 GHz
decreases as a function of radius and becomes $\tau_{\rm 32.96 GHz} \sim 1$ at $\theta \sim$ \msec{0}{032}. 
The emission of the optically thick inner region is $\sim 13$ mJy,
close to the observed flux. The total emission of the static 
atmosphere up to $r_g\sim130$ AU is $\sim 35$ mJy, and
we assume that the accretion disk has been 
truncated by the photoevaporation process beyond $r_g$.
Furthermore, as shown in Figure 5 of Lugo \& Lizano 
(2004), who modeled the photoevaporated disk wind in
the source MWC 349A, the static atmosphere
can produce a spectral index of the order of 1 for
frequencies between 20 and 100 GHz. 
On the other hand, according to H94, to preserve the static atmosphere
one requires a weak stellar wind because a strong
wind would blow the static atmosphere up to a radius $r_w$
where the ram pressure of the wind balances the 
thermal pressure of the ionized gas (see their Figure 1).
For the case of W3(OH), the condition that $r_w < r_g$,
imposes a condition on the stellar wind momentum
$M_{w-6} v_{w8} < 7.5 \times 10^{-2}$ (eq. [4.3] of H94),
where the stellar wind mass-loss rate $M_{w-6}$ is normalized to
$10^{-6}~ M_\odot~ {\rm yr}^{-1}$, and the stellar wind velocity $v_{w8}$
is normalized to 1000 km s$^{-1}$. Therefore, this condition implies 
a critical value for the mass-loss rate, $\dot M_{\rm crit}  \sim 10^{-7} M_\odot~ {\rm yr}^{-1}$,
which agrees with the value compiled by Dzib et al. (2013) for a B0 ZAMS star.

Finally, the flux variation on timescales of years, could
be explained by a variation in the wind strength: if the wind
momentum increases, it can blow away the static ionized
atmosphere decreasing the observed radio flux. When the star
returns to its low wind state, a static atmosphere will be regenerated
and the flux will increase to its high value. Variations in the wind strength could be
due to variations in the accretion rate through the disk as observed in young low-mass
stars (e.g., Pech et al. 2010).
In this scenario, sub-arcsecond sensitive millimetric observations
should detect the dust emission from the fossil circumstellar disk that is
being ionized by the central star.

\section{Conclusions}

In conclusion, the compact radio source projected near the center of W3(OH) coincides positionally
with a massive young star and has brightness temperature, spectral index, and polarization
characteristics suggestive of partially thick free-free emission. An interpretation
in terms of an ionized stellar winds fails because of the large flux densities
observed in the source. We tentatively propose that the
compact radio source could be the result of
either a slow shell ejection or the passage of a dense gas clump.
Finally, we discuss a scenario where the emission originates in a static 
ionized atmosphere around a fossil photoevaporated disk.

\acknowledgments

S.A.D., L.F.R., S.E.K., L.L., S.L. and L.A.Z. are thankful for the support
of DGAPA, UNAM, and of CONACyT (M\'exico).
This research has made use of the SIMBAD database, 
operated at CDS, Strasbourg, France.

\clearpage

\end{document}